\documentclass{emulateapj}
\usepackage{amsmath,amssymb}
\usepackage{graphicx,mathptm} 
\usepackage{times} 
\usepackage{natbib}
\usepackage{bm}
\usepackage{multirow}

\newcommand{\be}{\begin{equation}}
\newcommand{\ee}{\end{equation}}
\newcommand{\bea}{\begin{eqnarray}}
\newcommand{\eea}{\end{eqnarray}}

\shorttitle{Gyroresonance instability in MHD turbulence}
\shortauthors{Yan \& Lazarian}

\begin{document}
\title{Cosmic Ray transport through gyroresonance instability in compressible turbulence}

\author{Huirong
Yan\altaffilmark{1,2} \& A. Lazarian\altaffilmark{3}}
\altaffiltext{1}{ KIAA, Peking University, Beijing, 100871, China, hryan@pku.edu.cn}
\altaffiltext{2}{LPL, Steward Observatory, and Department of Physics, 
University of Arizona, Tucson, AZ 85721.}
\altaffiltext{3}{Department of Astronomy,
University of Wisconsin, Madison, WI 53706, alazarian@wisc.edu}

\label{firstpage}

\begin{abstract}
We study the nonlinear growth of kinetic gyroresonance instability of cosmic rays (CRs) induced by large scale compressible turbulence. This feedback of cosmic rays on turbulence was shown to induce an important scattering mechanism in addition to direct interaction with the compressible turbulence. The linear growth is bound to saturate due to the wave-particle interactions. By balancing increase of CR anisotropy via the large scale compression and its decrease via the wave-particle scattering, we find the steady state solutions. The nonlinear suppression due to the wave-particle scattering limit the energy range of CRs that can excite the instabilities and be scattered by the induced slab waves. The direct interaction with large scale compressible modes still appears to be the dominant mechanism for isotropization of high energy cosmic rays ($>100$ GeV).      
\end{abstract}


\section{introduction}

Cosmic rays and turbulence are essential components in astrophysical systems. Their interactions are crucial for high energy phenomena and dynamics of various systems.
The resonant and non-resonant (e.g., transient time damping, or TTD) interactions of cosmic rays with MHD turbulence are the accepted principal mechanism to scatter and isotropize cosmic rays \citep[see][]{Schlickeiser02}. In addition, efficient scattering is essential for the acceleration of cosmic rays. 
For instance, scattering of cosmic rays back into the shock is a vital component of the first order Fermi acceleration \citep[see][]{Longairbook}. At the same time, stochastic acceleration by turbulence is entirely based on scattering.

It is generally accepted that the energy of turbulence is most probably due to supernovae explosions\footnote{Stellar outflows are likely to be subdominant, while at the galactic outer regions magneto-rotational instability can play a role. The characteristic scale of the injection by those mechanisms is from dozens to hundreds of pc.} and cascaded down to small scales, where resonance with CRs of moderate energies happen\footnote{For very high energy CRs $\ga 100$TeVs, their resonant scales are close to the injection scale of turbulence $\ga $pc.}.

For years magnetic turbulence has been treated in ad hoc manner, with variations which included plane Alfv\'en waves moving parallel to magnetic field, a combination of these waves and so-called 2D perturbations, as isotropic magnetic turbulence with Kolmogorov spectrum and anisotropic turbulence with constant degree of anisotropy. Numerical simulation contradicted to all the models above, however.

Unlike hydrodynamic turbulence, Alfv\'{e}nic turbulence develops a scale-dependent anisotropic, with eddies elongated along the magnetic field \citep[][henceforth GS95]{GS95}. The GS95 model describes \textit{incompressible} Alfv\'{e}nic turbulence, which formally means that plasma $\beta\equiv P_{gas}/P_{mag}$, the ratio of gas pressure to magnetic pressure,s is infinity. But it may be conjectured that GS95 scaling should be approximately true for moderately compressible plasma \citep{LG01}. Studies in \cite[][henceforth CL02]{CL02_PRL} showed that the coupling of Alfv\'{e}nic and compressible modes is weak and that the Alfv\'{e}nic modes follow the GS95 spectrum in compressible medium. This is consistent with the analysis of observational data \citep[]{LP00,SL01}.

The turbulence injected on large scales may correspond to GS95 model and its extensions to compressible medium is less efficient in scattering of CRs compared to the estimates made assuming that magnetic turbulence consists of plane waves moving parallel to magnetic field. A cascade of Alfvenic perturbations initiated at large injection scales is shown to be really inefficient for the CR scattering \citep[]{Chandran00, YL02}. 


At the same time, one should not disregard the possibilities of generation of additional perturbations by CR themselves. For instance,  the slab Alfv\'enic perturbation can be created, e.g., via streaming instability \citep[see][]{Wentzel74, Cesarsky80}. These perturbations are present for a range of CRs energies (e.g., $\lesssim 100$GeV in interstellar medium for the streaming instability) owing to non-linear damping 
arising from ambient turbulence (\citealp[][henceforth YL02, YL04]{YL02, YL04}, \citealp{FG04}).
Instabilities induced by anisotropic distribution of CRs were also suggested as a possibility to scatter CRs \citep[]{Lerche, Melrose74}. 

Recent work by \citet[][henceforth LB06]{LB06} proposed that compressible turbulence can induce gyroresonance instability, which is an important feedback processes that can create slab modes to efficiently scatter CRs. This process is claimed to be more efficient in scattering CRs compared to direct action of turbulent fluctuations.

This gyro-kinetic instability is induced by the anisotropic distribution of CRs, which is caused by the compression arising from large scale turbulence. The degree of anisotropy, is determined by the compression on the scale of CR mean free path in LB06.  In their treatment the growth of CRs needs to be balanced with steepening in order to prevent the diffusion from approaching the Bohm limit, where the mean free path of particles becomes comparable to the Larmor radius. In this paper, we shall investigate the nonlinear suppression of the instability by considering the self-adjustment of anisotropy due to scattering with the magnetic perturbations. The instability reaches a stabilized growth rate due to the feedback of the increased perturbations on the anisotropy of CRs.

In what follows, in \S2 we give a brief background on turbulence and CR scattering, in \S3 we introduce the gyroresonance instability of CRs, in \S 4 we formulate our approach to the problem. We outline our findings and the implications of our results in \S5. In \S6 we discuss our results. The summary is provided in \S7. 

\section{MHD turbulence and cosmic ray scattering}

The propagation of cosmic rays (CRs) is mainly affected by their interaction with magnetic field. The properties of turbulence are vital for the correct description of CR propagation. For instance, the scattering frequency is much dependent on the anisotropy of turbulence \citep{LercheSchlickeiser, YL02}. 


Numerical simulations (\citealp*[see][]{CV00, MG01,MullerBisk,CLV_incomp,CL02_PRL,CL03} \citealp[see also the book of ][]{Biskampbook}, as well as, \citealp[][for reviews]{CLV_lecnotes, ElmegreenScalo}), show that Alfv\'enic modes exhibit scale-dependent anisotropy. This anisotropy was first discussed in \citet[][hencefoth GS95]{GS95}. However, the understanding came later that this anisotropy is observable only in the system of reference related to the {\it local} direction of magnetic field on the scale of the eddy under consideration (\citealp*[][]{LV99, CV00, MG01}). On the intuitive level it can be explained as the result of the following fact: it is easier to mix the magnetic field lines perpendicular to the direction of the magnetic field rather than to bend them. 
However, one cannot do mixing in the perpendicular direction to very small scales without affecting the parallel scales. This is probably the major difference between the adopted model of Alfv\'enic perturbations and the Reduced MHD \citep[see][]{Bieber94}. The corresponding
scaling can be easily obtained from the critical balance condition, i.e., $k_{\bot}\delta v_{k}\sim k_{\parallel}v_{A}$,
where $k_{\parallel}$ and $k_\bot$ are the parallel and perpendicular components of the wave vector
$\mathbf{k}$, $v_{A}$ is the Alfv\'en speed, $\delta v_k \propto k_\bot^{-\nu}$ is the turbulence velocity. Throughout the paper, $\|$ and $\bot$ are defined in reference to the local magnetic field.  As proven later by \citet{CLV_incomp}, the mixing motion perpendicular to magnetic field is essentially hydrodynamic  and energy in the turbulence cascade is conserved locally in phase space. From these arguments,
the scale dependent anisotropy $k_{\parallel}\propto k_{\perp}^{2/3}$ and $\delta v_k \propto k_\bot^{-1/3}\propto k_\|^{-1/2}$ can be derived.

Owing to the existence of the parallel cascade, Alfv\'enic turbulence can cascade to the resonant scale of CRs, which is characterized by the parallel scale $k_{\|,res}\sim \Omega/v$, where $\Omega$ is the Larmor frequency and $v$ is the particle speed. The scattering efficiency is, however, substantially reduced because of the anisotropy of the Alfv\'enic turbulence so that we can completely neglect it \citep{YL02,YL04} except for ultra-high energy CRs.  

The distribution of energy between compressible and incompressible modes depends, in general, on the way turbulence is driven. 
Naturally a more systematic study of different types of driving is required to determine the energy partition. In the absence of this, in what follows we assume that 
equal amounts of energy are transferred into fast and Alfv\'en modes when driving is at large scales.

While particular aspects of the GS95 model, e.g., the particular value of the spectral index, are the subject of controversies \citep[see][]{MullerBisk, Boldyrev, Gogo, BL09}, we think that, at present, GS95 model provides a good starting for developing models of CR scattering. MHD turbulence can be decomposed into Alfv\'en, slow and fast modes \citep{ CL02_PRL, CL03}. Slow modes are passive and follow the same GS95 scaling as do the Alfv\'en modes \citep{LG01, CL02_PRL}. Fast modes, instead, are marginally affected by Alfv\'en modes and follow acoustic type cascade \citep{CL03}. They have been identified by YL02,YL04, \citet[][hencefoth YL08]{YL08} as the major source of CR scattering in interstellar and intracluster medium.

Until recently, test particle approximation was assumed in most of earlier studies and no feedback of CRs is included apart from the streaming instability. Turbulence cascade is established from large scales and no feedback of CRs is included. This may not reflect the reality as we know the energy of CRs is comparable to that in turbulence and magnetic field \citep[see][]{Kulsrudbook}. It was suggested by LB06 that the gyroresonance instability of CRs can drain energy from the large scale turbulence and cause instability on small scales by the turbulence compression induced anisotropy on CRs. And the wave generated on the scales, in turn, provides additional scattering to CRs. In this paper, we would like to provide quantitative studies based on the nonlinear theory of the growth of the instability to take into account more accurately the feedback of the growing waves on the distributions of CRs.





\section{Gyroresonance instability of cosmic rays}

Plasma is known to be subject to numerous instabilities, one of them being gyroresonance instability \citep{Gary97_upper}. LB06 suggested that an analogous instability operates in CRs and is important for CRs subject to compressions arising from MHD turbulence.

\subsection{Origin of growth}
In a nonuniform plasma, where magnetic field strength varies, instability can develop due to anisotropy of particles arising from the adiabatic invariant $\mu$, which is preserved when the variations happen on length-scale much larger than  the particles' Larmor radii and timescale much longer than the Larmor period. 

To understand the essence of the instability it is important to recall that the definition for the magnetic moment of a charged particle in magnetic field is 
\be
\mu\equiv \frac{qv_\bot r_L}{2c}=\frac{p_\bot v_\bot}{2B},
\ee
where $q, p, r_L$ are the charge, momentum and Larmor radius of the particle, $c$ is the light speed, $B$ is the strength of the magnetic field. For non-relativistic particles, $\mu={\mathcal E}_{k,\bot}/B$ is proportional to the kinetic energy of particles ${\mathcal E}_{k,\bot}$; for relativistic particles, however, $p_\bot v_\bot$ deviates from the real kinetic energy. $\mu$ is an adiabatic invariant, indicating that $p_\bot v_\bot$ changes with the strength of magnetic field. In the regions where magnetic field increases, the perpendicular pressure of the particles is higher than the parallel one, inducing mirror and gyroresonance instability; in the places where magnetic field decreases, the opposite is true, resulting in firehose instability. Waves are generated through the instabilities, enhancing the scattering rates of the particles, their distribution will be relaxed to the state of marginal state of instability even in the collisionless environment. We define the degree of anisotropy of particle distribution as
\be
A\equiv\frac{W_\bot}{W_\|}-1,
\ee
where 
\bea
W_\bot&=&\int d^3pf({\bf p}) {\bf v_\bot p_\bot},\nonumber\\
W_\|&=&\int d^3pf({\bf p}) {\bf v_\| p_\|}
\eea
 is equivalent to the perpendicular and parallel components of the energy density of CRs.

In the presence of background compressible turbulence, the CR distribution is bound to be anisotropic because of the conservation of the adiabatic invariant $\mu$. Such anisotropic distribution is subjected to various instabilities.  While the hydrodynamic instability requires certain threshold, the kinetic instability can grow very fast with small deviations from isotropy. In this paper, we shall focus on the gyroresonance instability. The rate at which the anisotropy $A$ varies is determined by turbulence compressional rate, 

\bea
\left(\frac{\partial A}{\partial t}\right)_{gr}= \frac{1}{B}\frac{dB}{dt}=\omega \left(\frac{\delta v_\bot}{v_A}\right)
\label{Agrowth}
\eea
where $\omega\equiv v_{ph}/l_c$ is the frequency of the compressional modes at the compressional scale $l_c$, $\delta v_\bot \propto l_c^{\nu}$ is the perpendicular component of the turbulence velocity. 
Since beyond mean free path, the anisotropy is effectively reduced because of scattering, only compression $\lesssim\lambda $, the mean free path of CRs, can alter the degree of anisotropy. As $\omega \delta v_\bot/v_A \propto l_c^{\nu-1}$ and the power law index of turbulence $\nu$ is less than 1, the compression rate in fact increases with the decrease of the scale. So different from LB06, we adopt the turbulence damping scale  $l_c$ as the compression scale provided that $l_c\gg r_L$. For Kolmogorov scaling, $\nu=1/3$.  In the case of $\lambda<l_c$, as in some collisionless medium, the compression is still supplied from $l_c$, but with a reduced rate $ \omega \delta v_\bot/v_A \propto l_c^{-1+\nu}\lambda/l_c$. Using the compressional time $\omega^{-1}$ as the timescale for growth of the anisotropy, LB06 estimated the anisotropy $A\sim \delta v/v_A$. We notice, however, that the growth time for the anisotropy should be constrained by the scattering time of CRs $\tau_{scatt}$. If $\tau_{scatt}\simeq \lambda/c$ were adopted,  the anisotropy of CRs $A\sim \tau_{scatt} \partial A/\partial t$ would not reach the threshold $v_A/v$ \citep[see][]{Kulsrudbook} unless the mean free path of CRs is much larger than the turbulence compressional scale, which is limited by the damping%
\footnote{In collisionless environments, for instance, the dissipation scale of compressible modes can be very large \citep{YL04, Brunetti_Laz}}. In fact, this constraint does not apply because we are dealing with an evolving system rather than an equilibrium one. The wave amplitude $\epsilon$ is growing at a rate $\Gamma_{gr}$ and so the mean free path is decreasing. Therefore the  scattering rate is decreased compared to the equilibrium value $\Omega/\epsilon_N\equiv \Omega B^2/8\pi\epsilon$. The reduced rate may be estimated as below. The variation of anisotropy A can be expressed as 
\be
\frac{d A}{dt}\simeq \nu A=\frac{1}{W_\|}\left(\frac{dW_\bot}{dt}-\frac{W_\bot}{W_\|}\frac{dW_\bot}{dt}\right)\sim \Gamma_{gr}\epsilon_N/\beta_{CR},
\ee 
where $\beta_{CR}\equiv 8\pi W_\|/B^2$ is the ratio of CR pressure to magnetic pressure. In the above equation, we utilized the fact that $A<<1$ and so $W_\bot/W_\|\simeq 1$ and that during the isotropization induced by the growing waves, the loss in $W_\bot$ is greater than the gain in $W_\|$ so that wave can grow at the expense of the CRs' energy. From the above equation, the isotropization rate can be estimated as

\be
\tau^{-1}_{scatt} \sim   \frac{\Gamma_{gr}\epsilon_N}{\beta_{CR} A}.
\label{nu_est}
\ee 
By balancing the rate of decrease in anisotropy due to scattering and the growth due to compression, we get
\bea
\epsilon_N&\sim& \frac{ \beta_{CR}\omega\delta v}{\Gamma_{gr} v_A },\nonumber\\
\lambda&=&r_p/\epsilon_N.
\label{epsilon_est}
\eea

The magnetic field strength is altered by compressible modes. In low $\beta$ medium, the compression is mainly from fast modes. For fast modes, their cascade is slower than their wave oscillation. This means that they have periodical oscillations unlike slow modes turbulence. A compression/expansion is followed by another, therefore a growth rate must be faster than the wave frequency in order for the instability to grow \citep{Yan09}. 
In high $\beta$ medium,  the situation is different and the contribution from slow modes is dominant. 

In the above equations, we did not account for collisions. This is physical because the scale we consider is less than the mean free path.
In LB06, the only feedback was passively incorporated in $\lambda$, which decreases as the wave grows. This process, however, will not stop until $\lambda$ reaches Larmor radius, namely, the Bohm diffusion regime. 
In this paper, we shall provide a quantitative treatment of the nonlinear suppression.



\subsection{Growth rate}

The distribution function of CRs $f({\bf x, p},t)$ is governed by the Vlasov equation \citep[see, e.g.][]{Kulsrudbook}:
\be
\frac{\partial f}{\partial t}+{\bf v\cdot}\frac{\partial f}{\partial {\bf x}}+q\left[{\bf E(x},t)+\frac{{\bf v\times B}}{c}\right]\cdot \frac{\partial f}{\partial {\bf p}}=0,
\ee
where ${\bf v, p}$ are velocity and the momentum of the particles respectively. ${\bf E}$ is the electric field.  We can take a perturbation expansion for the distribution function 
\be
f=f_0+f_1+f_2+..., \,~~~~~ f_{i+1}\ll f_i
\ee
if the field perturbation is small $\delta B\ll B_0$, where $B_0$ is taken in ${\hat z}$ direction\footnote{The gyroradii are much smaller than the turbulence injection scale unless one is considering ultra-high energy CRs. So the condition $\delta B\ll B_0$ is well justified for moderate energy CRs.}. 

Assuming a monochromic Alfv\'en wave propagating along z direction,  the linear perturbation $f_1$ can be obtained using the quasi-linear theory \citep[see e.g.][]{Kulsrudbook}.  The linearization of the Vlasov equation is

\be
-i\omega f_1+ikv_z f_1+ \frac{q}{c}({\bf v \times B_0}) \cdot \nabla_p f_1= -q\left[{\bf E + \frac{v\times (k \times E)}{\omega}}\right]\cdot\nabla_p f_0
\ee
By integrating it, one can get

\be
f_1^{\pm}=\frac{iqv_\pm E_\mp}{\omega}\left(k\frac{\partial f_0}{\partial p_\|}+\frac{\omega-k_\|v_\|}{v_\bot}\frac{\partial f_0}{\partial p_\bot}\right)(k_\|v_\|-\omega\mp\Omega)^{-1},
\label{linearf}
\ee
where $f_1^+$ corresponds to the right-moving cosmic rays and $f_1^-$ is for left-moving cosmic rays. $\omega, k$ are the wave frequency and the wave number. $\Omega=\Omega_0/\gamma_L$ is the relativistic Larmor frequency, $\Omega_0=qB/m/c$, $m$ is the proton mass. From the Maxwell equations, one can obtain
\be
{\bf E}_1\left(1-\frac{k^2c^2}{\omega^2}\right)+\frac{4\pi i {\bf J}_1}{\omega}=0,
\label{maxwell}
\ee
where 
\be
{\bf J}_1=\sum q\int d^3v {\bf v}f_1
\label{current}
\ee
is the current density arising from the perturbed distribution $f_1$.
Then,  the dispersion relation can be obtained through eqs.(\ref{linearf},\ref{maxwell},\ref{current}) 
\be
1-\frac{k^2c^2}{\omega^2}+\sum_j K_j^{\pm}({\bf k},\omega)=0
\label{dispersion}
\ee
and the perturbed part of dielectric tensor
\be
K_j^\pm\equiv \frac{4\pi i {\bf J}_1}{\omega {\bf E}_1} =\frac{2\pi q^2}{\omega^2}\int d^3p \frac{v_\bot}{\omega-k_\|v_\|\pm\Omega} \left[kv_\bot\frac{\partial f_0}{\partial p_\|}+(\omega-k_\|v_\|)\frac{\partial f_0}{\partial p_\bot}\right],
\label{dielectric}
\ee
where the subscript `j' refers to different species.
For $\omega\ll \Omega$, $K^0=c^2/v_A^2$, representing MHD waves parallel to the magnetic field with $\omega_0=kv_A$. Insert $\omega=\omega_0+\omega_1$ into the above equation, one can find the growth rate corresponding to the imaginary part of the perturbed $K_j$,
\be
\Gamma^\pm=\Im(\omega_1)=-\frac{\omega_0v_A^2}{2c^2}\Im(K_j^\pm)
\label{growth_def}
\ee    

The growth rate of the gyroresonance instability can be easily derived from  this equation:
\bea
\Gamma^{\pm}_{gr}&=&\pi^2e^2v_A\int \frac{v_\bot^2}{c^2}\left(\frac{\partial f}{\partial p_\|}-\frac{v_\|}{v_\bot}\frac{\partial f}{\partial p_\bot}\right)\delta (k_\|v_\|\mp \Omega)d^3{\bf p}\nonumber\\
&=&\omega_p\frac{n_{cr}(p>m\Omega_0/k_\|)}{n}(A-A_{cr})Q, \nonumber\\
&=&\frac{v_A}{L_i}(A-A_{cr})\left(\frac{r_p}{r_0}\right)^{1-\alpha},\nonumber\\
\label{growth}
\eea
where $n$ is the gas number density, $\omega_p=\sqrt{4\pi n e^2/m}$ is the proton plasma frequency, $n_{cr}\propto p^{1-\alpha}$ is the number density of cosmic rays, and (see e.g.,  LB06)
\bea
A_{cr}&=&v_A/v,\nonumber\\
Q&=& \frac{\pi}{16}(\alpha+2)(\alpha-1) Bt(3/2,\alpha/2),\nonumber\\
L_i&=&6.4\times 10^{-7}\left(\frac{B}{5\mu {\rm G}}\right)\left[\frac{4\times 10^{-10}{\rm cm}^{-3}}{n_{cr}(r_p>r_0)}\right]{\rm pc},
\label{main}
\eea
in which $r_0$ is the gyro-radius of 1GeV particles, $Bt$ represents a beta function \citep[see][]{Math_handbook}.

\begin{figure}
\includegraphics[width=0.95\columnwidth,
  height=0.24\textheight]{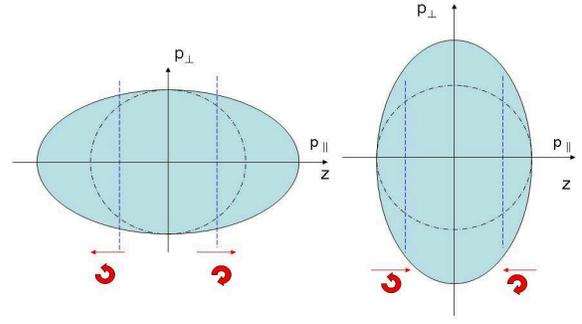}
 \caption{Anisotropy of particle distributions and waves excited by them. The oblate distribution on the left can be caused by rarefaction/expansion of the field and the prolate on the right can be realized due to compression of the field. The dashed lines represent the resonant momenta $p_{res}$. The dashdot circles are the paths along which the derivative of f needs to be considered (see Eq.\ref{fdot}). As explained in the text, for each distribution, right moving and left moving CRs excite right-handed and left-handed polarized (annular arrows) waves moving in opposite directions (the lines with arrows). }
 \label{vel_dist}
\end{figure}

The sign of $\Gamma_{gr}$ determines whether the wave grows ('+') or damps ('-'). It depends on the bracketed expression in Eq.(\ref{dielectric}), which is a derivative of $f$ along a circle centered at $p_z = m v_A, p_\bot = 0$ \citep[see][]{Kulsrudbook}, 
\be
\frac{d p_\bot}{d p_\|}=\frac{\omega-kv_z}{kv_\bot}.
\label{fdot}
\ee
The wave grows if there are more and more particles along the way it propagates. Unlike with shifted distribution (streaming instability), only circularly polarized waves can be excited by the gyroresonance instability. First we decide the directions of wave vector resonant with right and left moving cosmic rays, respectively. Then we determine the polarization of the waves bearing in mind that CR protons always rotate clockwise with respect to the magnetic field. For an oblate distribution of particles with $p_\bot>p_\|$, the left moving, right circularly polarized wave can resonate with the right moving CRs (with superscript `+') and gain energy from them as it encounters more particles in the wave propagation direction (see Fig.\ref{vel_dist}{\it right}). So do the right moving, left circularly polarized waves with the left moving CRs (with superscript `-'). For a  prolate distribution, the opposite is true. They are illustrated by Fig.\ref{vel_dist}.

\section{Feedback processes and steady state growth solutions}
\subsection{Nonlinear suppression}

In this section we shall provide a rigorous study of the suppression of the growth of the instability. The wave-particle interaction was treated using the second order theory (SOT) in \citet{GaryTokar} for bi-Maxwellian distribution of thermal particles. These particles have Maxwellian distribution characterized by $T_\|, T_\bot$, two different temperatures parallel and perpendicular to the local magnetic field.  We shall modify the SOT for the CRs' power-law distribution, i.e. $f(p)\propto p^{-\alpha-2}$.  

Assuming small-amplitude fluctuation, we can easily extend the Vlasov equation to the second order. The slowly varying second order distribution of particles due to the stochastic interactions with the electro-magnetic perturbations can be described by the ensemble average of the second order extension of the Vlasov equation, 

\bea
&&\frac{\partial <f_2>}{\partial t}+\frac{q{\bf v}\times {\bf B}}{c}\cdot\frac{\partial<f_2>}{\partial {\bf p}}+q{\bf E}_2\cdot\frac{\partial f_0}{\partial {\bf p}}\nonumber\\
&=&-e<\left({\bf E}_1+\frac{{\bf v}\times {\bf B}_1}{c}\right)\cdot \frac{\partial f_1}{\partial {\bf p}}>,
\label{2nddist}
\eea
The evolution of the anisotropy of the CRs $A$ can be derived from this equation.  The electric field ${\bf E}_1$ is perpendicular to the magnetic field and therefore increases the perpendicular energy of CRs. The magnetic perturbation ${\bf B}_1$ scatter particles and thus only causes exchange of particle energies in the parallel and perpendicular direction. The second moment of Eq.(\ref{2nddist}) gives

\bea
\frac{\partial W_\bot}{\partial t}&=&2q<{\bf E}_1\cdot \int {\bf v}f_1 d^3p>+\frac{2q}{c}<{\bf B}_1\cdot \int d^3p {\bf v}_\bot\times {\bf v}f_1>,\nonumber\\
\frac{\partial W_\|}{\partial t}&=&-\frac{2q}{c}<{\bf B}_1\cdot \int d^3p {\bf v}_\bot\times {\bf v}f_1>,
\label{2ndmoment}
\eea
The two terms on the right hand side of the first equation correspond to the acceleration and scattering. They are given by (see Appendix for the detailed derivation),
\bea
&&q<{\bf E}_1\cdot \int {\bf v}f d^3p>=\int dk\frac{2}{|\omega|^2}\Im\left[\omega^2\omega^*(K^+\epsilon_-+K^-\epsilon_+)\right],\nonumber\\
&&\frac{q}{c}<{\bf B}_1\cdot \int d^3p {\bf v}_\bot\times {\bf v}f_1>=-\int dk\frac{2}{|\omega|^2}
\left\{\omega_{pr}^2\left(\Gamma^+\epsilon_-+\Gamma^-\epsilon_+\right)\right.\nonumber\\
&+&\left.\Im\left[(\omega_r+\Omega)\omega^2K^+\epsilon_-+(\omega_r-\Omega)\omega^2K^-\epsilon_+\right]\right\},
\label{forces}
\eea
where $\omega_{pr}\equiv \sqrt{4\pi n_{cr}(p>p_{res})q^2/m}$ is equivalent of plasma frequency, but for CRs, $\omega_r$ is the real part of $\omega$.

\be
\epsilon_\pm\equiv \frac{E^*_\pm E_\pm}{8\pi},
\ee

For a symmetric distribution, we expect that $\epsilon_+=\epsilon_-$ and $\Gamma^+=\Gamma^-$. Then from Eqs.(\ref{2ndmoment},\ref{forces}), we can get

\bea
\frac{\partial W_\bot}{\partial t}&=&\int dk S_\bot(k)\frac{\partial \epsilon(k)}{\partial t},\nonumber\\
\frac{\partial W_\|}{\partial t}&=&\int dk S_\|(k)\frac{\partial \epsilon(k)}{\partial t},
\label{momentumrate_int}
\eea
where

\be
\epsilon(k)\equiv \frac{|B_1|^2+|E|^2}{8\pi}= \left(1+\frac{k^2c^2}{|\omega|^2}\right)(\epsilon_++\epsilon_-)
\ee
is the total energy density of the wave and $S_\bot, S_\|$ are measures of the scattering efficiency, determining how fast is the response of the distribution of CRs to the wave perturbations. They increase with the decrease of wave frequency (see Appendix for the detailed derivation). In general, their function is to reduce the anisotropy. 

For gyroresonance instability, the growth of the wave happens at the gyroresonance frequency of the CRs, $k_{res}$. Therefore in the first order approximation the above evolution equation can  be differentiated, and accordingly we obtain (see Appendix for detailed derivation)

\bea
\frac{\partial W_\bot(k)}{\partial t}&=&S_\bot(k)\frac{\partial \epsilon(k)}{\partial t},\nonumber\\
\frac{\partial W_\|(k)}{\partial t}&=&S_\|(k)\frac{\partial \epsilon(k)}{\partial t},
\label{momentumrate}
\eea
and
\bea
S_\bot(k)&=&-2\left[\frac{2(\alpha+2)}{(\alpha+1)(\alpha+3)}\omega_{pr}^2+2k^2c^2\right)/\left(|\omega|^2+c^2k^2\right],\nonumber\\
&\simeq&-2\left[\frac{(\alpha+2)}{(\alpha+1)(\alpha+3)}\beta_{CR}(p>p_{res})+2\right],\nonumber\\
S_\|(k)&=&2\left[\frac{\alpha(\alpha-1)}{(\alpha+2)}\omega_{pr}^2+k^2c^2\right)/\left(|\omega|^2+c^2k^2\right],\nonumber\\
&\simeq&2\left[\frac{(\alpha+2)}{(\alpha+1)(\alpha+3)}\beta_{CR}(p>p_{res})+1\right].
\label{Svalue}
\eea

From this, we can get
\bea
\frac{\partial A}{\partial t}=\partial \left(\frac{W_\bot}{W_\|}\right)/\partial t=\frac{1}{W_\|}\left[S_\bot-(A+1)S_\|\right]\epsilon \Gamma_{gr}
\label{Asuppr}
\eea
 
By equating Eqs.(\ref{Agrowth}),(\ref{Asuppr}), we get
\be
S_\bot \Gamma_{gr} \epsilon_N + \beta_{CR} \omega\left(\frac{\delta v_\bot}{v_A}\right)-(A+1)S_\|\Gamma_{gr} \epsilon_N=0, 
\label{general_solution}
\ee
Since $S_\|, S_\bot$ are of order 1 (see Fig.1), we see that the results from above equation is comparable to our earlier estimate in Eq.(\ref{epsilon_est}).
There are two unknowns $A, \epsilon_N$ in the equation (\ref{general_solution}). There has to be another relation to determine them.  We assume here that the system relaxes to the marginal state of instability, and $A=A_{cr}=v_A/v$. For the compression by fast modes, $\omega= v_A/l_c$., $\delta v_\bot \approx V(l_c/L)^{\nu}$, $V$ the injection velocity of turbulence at the injection scale $L$.  We get for the compression by fast modes,
\bea
\epsilon_N&=& \beta_{CR} M_A \left(\frac{l_c}{L}\right)^{\nu}\frac{L_i}{l_cA}\left(\frac{r_p}{r_0}\right)^{\alpha-1}\nonumber\\
&/&\left[S_\parallel (1+A)-S_\bot\right]\,, ~if ~\lambda>l_c
\label{lambdlslc}
\eea
\bea
\epsilon_N&=& \left\{\beta_{CR} M_A\frac{r_p}{l_c} \left(\frac{l_c}{L}\right)^{\nu}\frac{L_i}{l_cA}\left(\frac{r_p}{r_0}\right)^{\alpha-1}\right.\nonumber\\
&/&\left.\left[S_\parallel (1+A)-S_\bot \right]\right\}^{\frac{1}{2}}\,, ~if ~\lambda<l_c
\label{lambdgtrlc}
\eea
where $M_A\equiv V/v_A$ is the Alfv\'enic Mach number. $\nu=1/3$ for the Kolmogorov turbulence and $\nu=1/4$ for the Iroshinikov-Kraichnan turbulence. We adopt Kolmogorov scaling\footnote{At a given scale $l_c$, the compression and wave growth would be stronger if  IK scaling is used.} for the fast modes in this paper.

The parallel damping scale of slow modes in high $\beta$ medium is the proton diffusion scale $\beta^{1/2}l_{mfp}$, where $l_{mfp}$ is the mean free path of thermal particles (\citealt[]{Barnes66}).  For slow modes in high $\beta$ medium, $\omega= v_A/l_c$, $\delta v_\bot/v_A=\delta v/v_A \cos\theta\sim (l_c/ l_A)^{-1/2}$, where $l_A=L/M_A^3$ in the case of $M_A>1$. Inserting it into Eq.(\ref{general_solution}), we get  

\bea
\epsilon_N&=&\beta_{CR} \left(\frac{l_c}{l_A}\right)^\frac{1}{2}\frac{L_i}{l_cA}\left(\frac{r_p}{r_0}\right)^{\alpha-1}\nonumber\\
&/&\left[S_\parallel (1+A)-S_\bot \right] \,, ~ if~\lambda>l_c
\label{slowmodes_gtr}
\eea
\bea
\epsilon_N&=&\left\{\beta_{CR} \frac{r_p}{l_c}\left(\frac{l_c}{l_A}\right)^\frac{1}{2}\frac{L_i}{l_cA}\left(\frac{r_p}{r_0}\right)^{\alpha-1}\right.\nonumber\\
&/&\left.\left[S_\parallel (1+A)-S_\bot \right]\right\}^\frac{1}{2}\,, ~ if ~\lambda<l_c
\label{slowmodes_ls}
\eea

\subsection{bottle-neck for the growth due to energy constraint}
\label{feedback}
The creation of the slab waves through the CR resonant instability is another channel to drain the energy of large scale turbulence. This process, on one hand, can damp the turbulence. On the other hand, it means that the growth rate is limited by the turbulence cascade. The energy growth rate $\Gamma_{gr}\epsilon $ cannot be larger than the turbulence energy cascading rate, which is $1/2 \rho V^4/v_A/L$ for fast modes in low $\beta$ medium and $\rho v_A^3/l_A$ for slow modes in high $\beta$ medium. This places a constraint on the growth:

\bea
\Gamma_{gr} \epsilon_N\le \begin{cases}M_A^4 v_A/L, & \beta<1,\\
  v_A/l_A, & \beta>1,\end{cases}
\eea  
Thus the upper limit of wave energy is given by

\bea
\epsilon^u_{N}=\begin{cases}M_A^2L_i/(L A)(r_p/r_0)^{\alpha-1},& \beta<1 \\
  L_i/(l_A A)(r_p/r_0)^{\alpha-1}, & \beta>1. \end{cases}
\label{energy}
\eea

The growth is induced by the compression at scales $\lesssim \lambda$. Therefore, in the case that $\Gamma_{gr} \epsilon_N$ reaches the energy cascading rate, fast modes are damped at the corresponding maximum turbulence pumping scale $\lambda_{fb}=r_p/\epsilon_N$.  If $\lambda_{fb}$ is larger than the original damping scale $l_c$, then $l_c$ should be replaced by $\lambda_{fb}$, and one needs to iterate until the results are self-consistent.


\begin{figure}
\includegraphics[width=0.95\columnwidth,
  height=0.24\textheight]{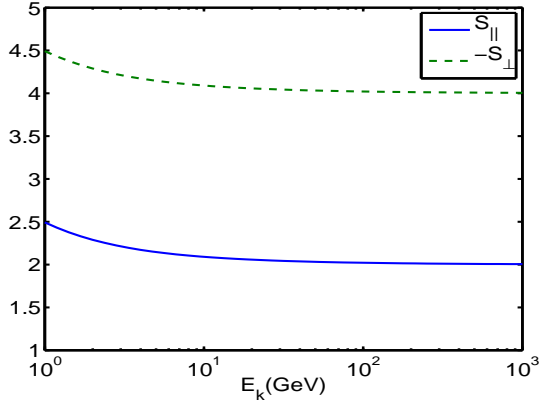}
\caption{Scattering coefficients $S_\bot, S_\|$ vs. CR energy.}
\label{Acomparison}
\end{figure}

\begin{figure}
\includegraphics[width=0.98\columnwidth,
  height=0.28\textheight]{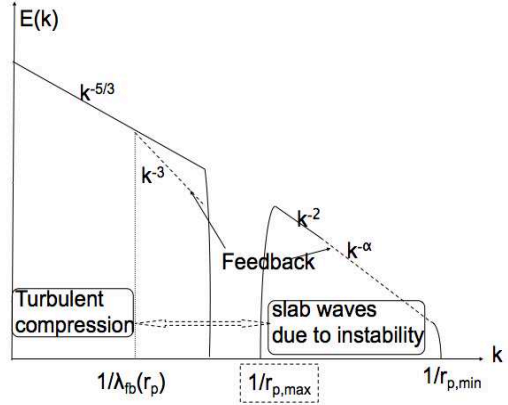}
\caption{The spectral energy density of slab waves that is transferred from the large scale compressible turbulence via the gyroresonance instability of CRs. In the case that the instability grows up to the maximum energy rate allowed by the turbulence cascade, large scale turbulence is truncated at $\lambda_{fb}$ and the wave amplitude $E(k)dk\sim \epsilon_N^u$ is given by Eq.(\ref{energy}). Note that the picture is different from LB06, namely, the feedback on the large scale turbulence occurs only in some cases when the scattering is not sufficient to prevent the waves from growing to the maximum values.}
\label{feedback}
\end{figure}


\subsection{Summary of Procedures}

The gyroresonance instability can be driven by CRs with anisotropic distribution which is induced by compressible turbulence. Fast and slow modes dominate the compression in low and high $\beta$ medium respectively. The compressional rate is maximized at the smallest scale-damping scale of the turbulence. Since CRs do not `remember' the perturbation of magnetic field and its anisotropy beyond its mean free path, compressions larger than the mean free path of CRs will not induce anisotropy. Therefore, we adopt min$(\lambda, l_c)$ as the compressional scale that drives the anisotropy of CRs. We first  calculate the damping scale of the compressible modes. If the damping scale is larger than the original mean free path $\lambda_0$, one should use the Eq.(\ref{lambdlslc},\ref{slowmodes_gtr}) to obtain the wave amplitude $\epsilon_N$  and the new mean free path $\lambda=r_p/\epsilon_N$. If the damping scale is smaller than the mean free path $\lambda_0$, we use Eqs.(\ref{lambdgtrlc},\ref{slowmodes_ls}) to get  $\epsilon_N$ and $\lambda=r_p/\epsilon_N$.  

The resulting energy growth rate $\Gamma_{gr}\epsilon$ should be compared with the large scale turbulence energy cascading rate $V^3/L$.  If it is larger than   $V^3/L$, the growth is limited by the energy budget, rather than the nonlinear feedback. Then eq.(\ref{energy}) can be used to estimate the growth (see Fig.\ref{feedback}).

 The growth rate should also overcome the damping by background turbulence (YL02, 04; Farmer \& Goldreich 2004, \citealp*[][]{BL08}). If the growth rate is lower than the damping rate
 \be
 \Gamma_d=V_{LM}/\sqrt{L_M r_p},
 \label{turbdamp}
 \ee
the instability does not operate. $L_M, V_{LM}$ are the injection scale of strong MHD turbulence. $L_M=l_A$, $V_{LM}=v_A$ if $M_A>1$, and $L_M=LM_A^2, V_{LM}=v_AM_A^2$ if $M_A<1$.

For fast modes, one also needs to take into account the turbulence compression rate at the dissipation scale $\omega=v_A/l_c$. If the growth rate is smaller than the compression, the subsequent expansion will smear out the anisotropy, the trigger of the kinetic instability, and the growth will not happen.

We show results of our study in figures \ref{halo}, \ref{HIM}, \ref{ICM}. In the plots, we also demonstrate several relevant dimensionless parameters: degree of anisotropy A and the ratio of CR pressure to magnetic pressure, $\beta_{CR}$, as well as the growth and $\omega$. The mean free path of CRs are plotted together with the earlier estimates from LB06. Compared to LB06, the degree of anisotropy A is much smaller, and set by the marginal state of instability. This results in a much lower growth rate, and therefore smaller range of energies of CRs that can induce the instability because of the constraint by damping by background turbulence. In addition, the nonlinear feedback due to the scattering by the growing waves is more efficient in saturating the growth at small amplitude ($\epsilon_N<1$) compared to the feedback considered in LB06. We do not need additional wave steepening to stop the wave from growing to large amplitude (or the diffusion to reach the Bohm limit). Our treatment is self-consistent. 


\section{Astrophysical Implications}

\begin{table*}
\begin{tabular}{ccccccccc}
\hline
\hline
&T(K)&n(cm$^{-3}$)&B($\mu$G)&$M_A$&L(pc)&$n_{cr}(10^{-10}{\rm cm}^{-3})$&$l_c$(cm)&$\lambda_{fb}$(cm)\\
\hline
halo&$2\times 10^6$&0.001&5&1&30&4&$2\times 10^{19}$&$1.84\times 10^{17}$\\
HIM&$ 10^6$&0.005&2.5&1&30&4&$4.9\times10^{16}$&--\\
ICM&$10^8$&$0.001$&1&10&$3\times 10^5$&4&$4.2\times 10^{19}$&--\\
\hline
\hline
\end{tabular}
\caption{The physical parameters we adopted for different medium. HIM=hot ionized medium, ICM=general intracluster medium.}
\end{table*}

\subsection{Importance of the instability}

Our present work confirms the finding in LB06 that the gyroresonance instability plays an important role for the propagation of cosmic rays in turbulent astrophysical fluids. We note, that the coupling of the compressible and incompressible motions reported in Cho \& Lazarian (2003) between compressible and Alfv\'enic modes is strong enough at the scale of driving for diverting about 20\% of energy into compressible cascade, even if the driving is purely solenoidal. While we still do not know many details of the MHD cascade (see Beresnyak \& Lazarian 2009), the fact that a noticeable portion of energy is going into large scale compressions is generally accepted. It is those compressions that the gyroresonance instability requires to induce substantial changes of the dynamics of the astrophysical fluids.

In view of the above understandings, it looks wrong that present day numerical codes ignore the effects of cosmic rays. We may expect an appreciable change of the compressible dynamics if cosmic rays drain energy from the compressible motions. The consequences of the process for the ISM and molecular cloud formation will be discussed elsewhere.

At the same time, the plane parallel Alfv\'en waves with $k_{\|}\gg k_{\bot}$ are expected to interact efficiently with cosmic rays\footnote{In the Solar and Magnetospheric community it is accepted to talk about the energetic particles rather than cosmic rays. This distinction is not important for the basic processes we discuss.}. Apart from scattering this should induce cosmic ray acceleration via second order Fermi process. The consequence of this important process is still to be evaluated quantitatively for different astrophysical environments.

Our quantitative study shows that LB06 in their simplified treatment overestimated the growth rate and therefore the range of energies of cosmic rays over which the instability is active. The reason is that the degree of anisotropy A is overestimated in LB06. The nonlinear feedback that we study in this paper limit the growth through the scattering and therefore A is maintained at the level of marginal instability.  Due to the same scattering, our steady state wave energy/mean free path is smaller/larger than that in LB06 (see the right panels of Fig.\ref{halo}, \ref{HIM}, \ref{ICM}). The quenching of growth they considered by only reducing the mean free path is not adequate, and in fact does not apply here as we consider the compression at the minimum scale of turbulence instead of the mean free path.  
Turbulence energy cascading rate is another constraint that sometimes is more stringent than the scattering \footnote{The effect on turbulence cascade was considered in LB06, but not the feedback on the wave growth.}. 
 
\subsection{Particular cases of astrophysical environments}

\subsubsection{Galaxy}
\label{galaxy}
\begin{figure*}
\includegraphics[width=0.24\textwidth,
  height=0.18\textheight]{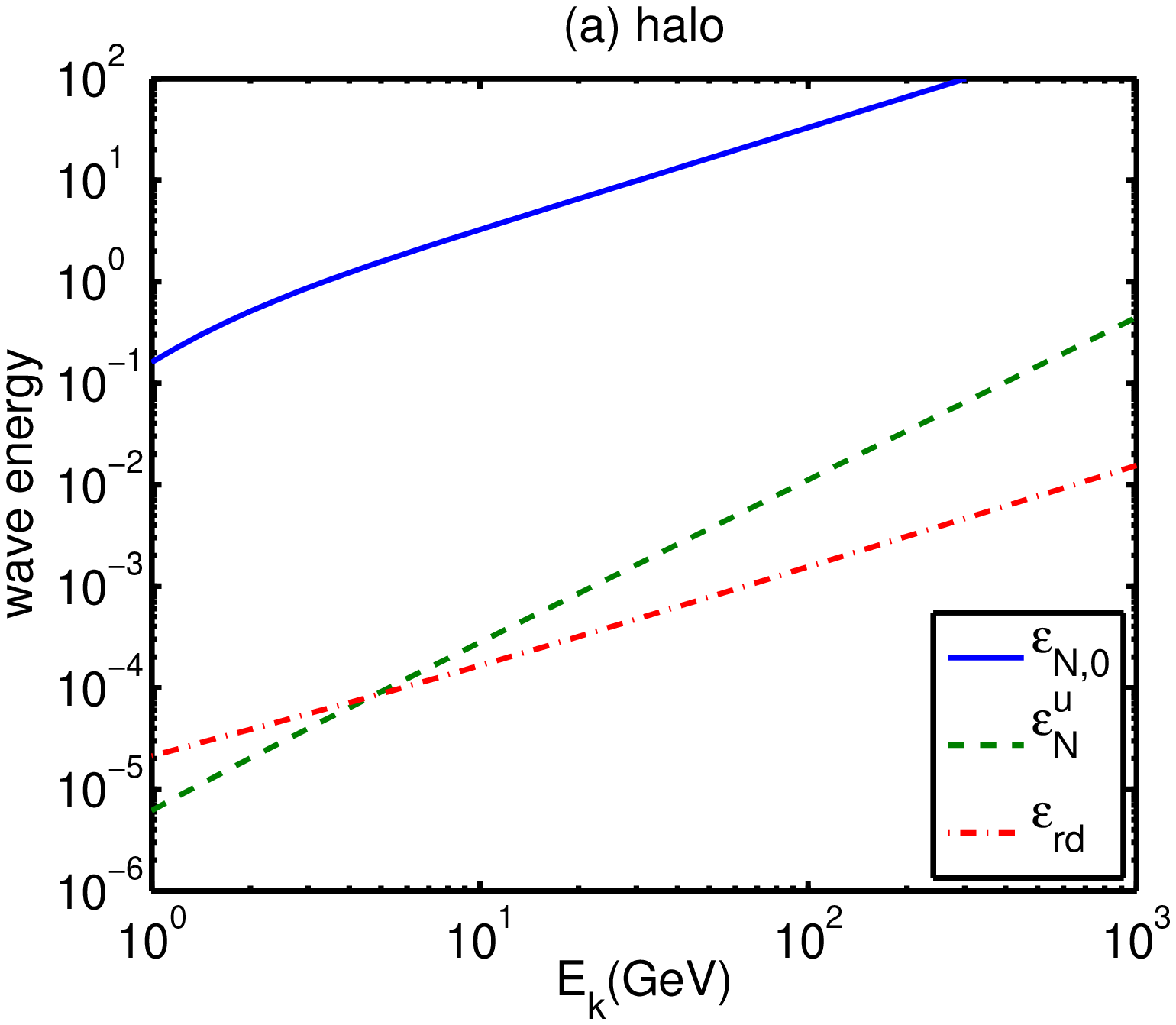}
\includegraphics[width=0.24\textwidth,
  height=0.18\textheight]{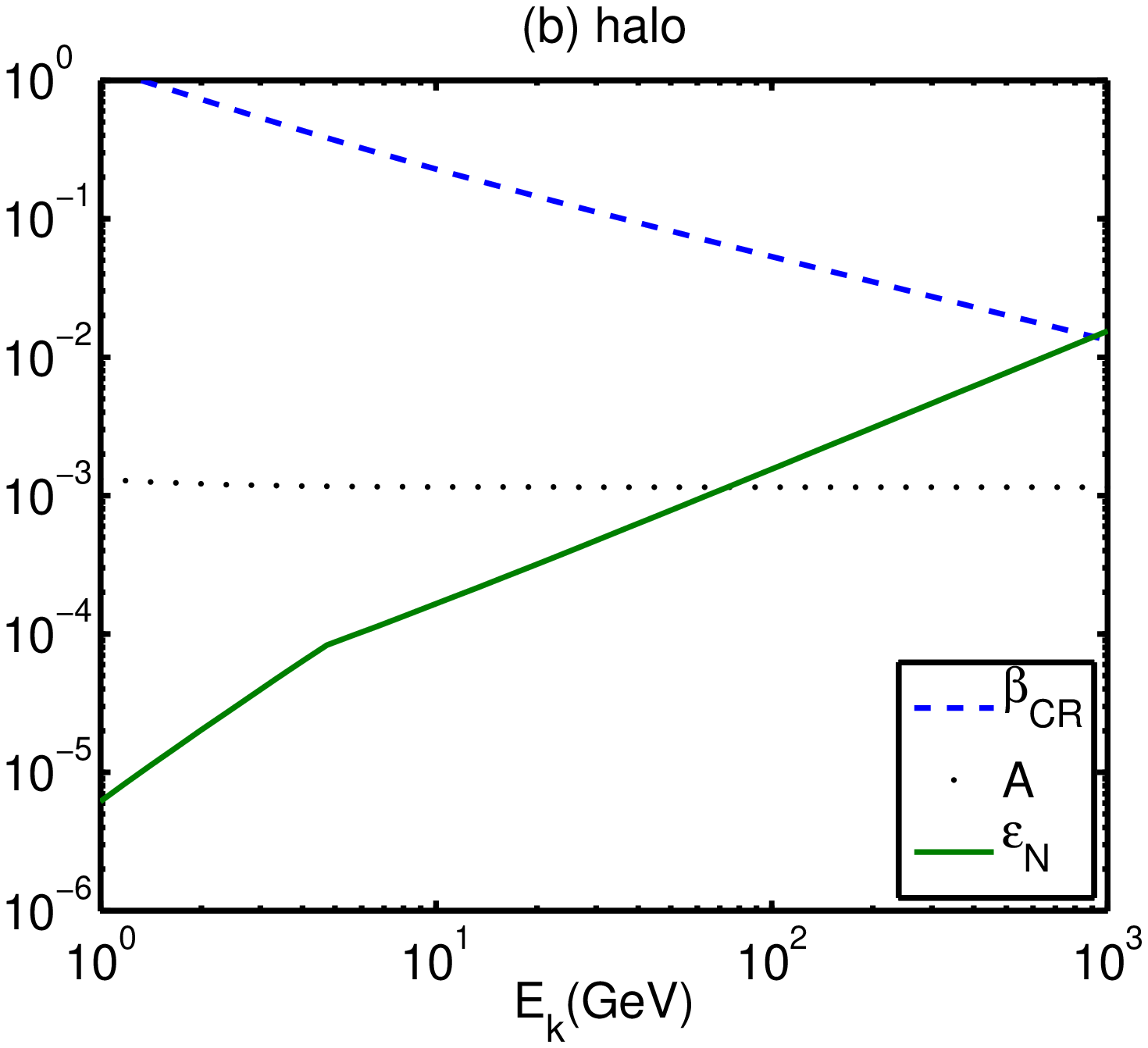}
  \includegraphics[width=0.24\textwidth,
  height=0.18\textheight]{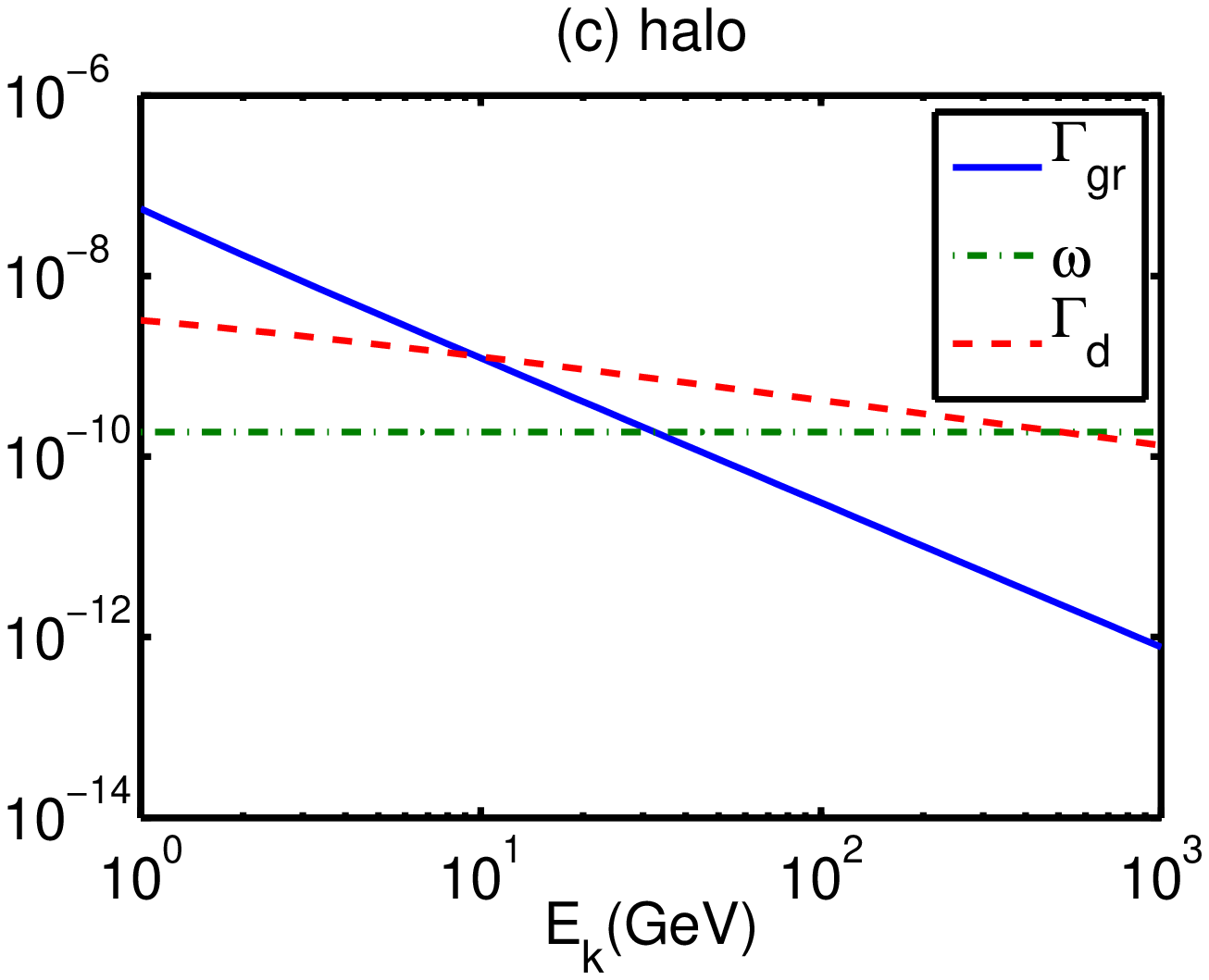}
   \includegraphics[width=0.24\textwidth,
  height=0.18\textheight]{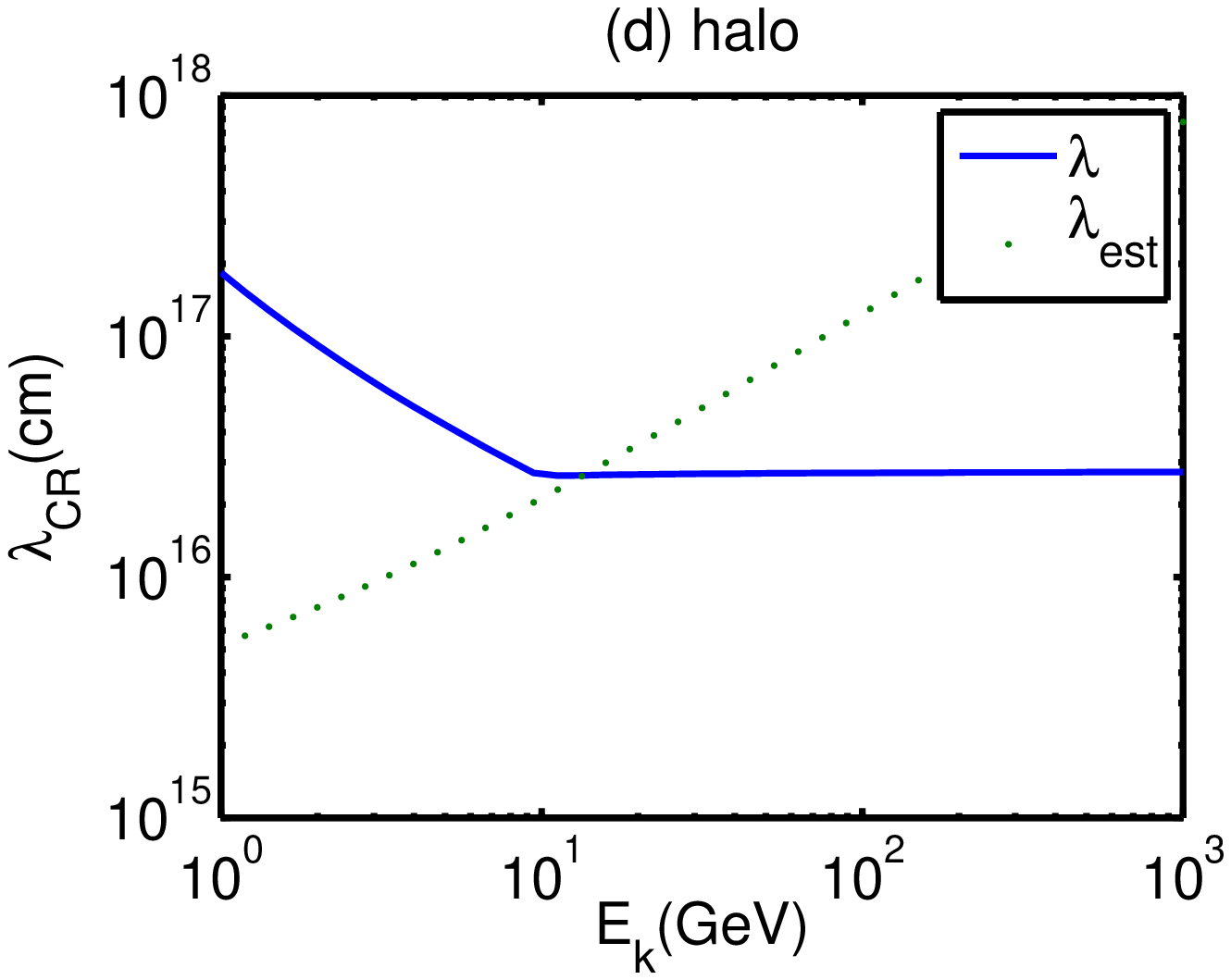}
  \caption{(a): the amplitude of the growing wave using different approaches. $\epsilon_{N,0}$ is obtained with pumping at the mean free path, which exceeds the maximum value $\epsilon^u_{N}$ allowed by the energy of turbulence and is therefore unrealistic;  $\epsilon_{rd}$ is calculated using the reduced compression from the damping scale set by the maximum mean free path obtained through $\epsilon_{N,0}$; the real amplitude is the minimum of $\epsilon_N={\rm min}(\epsilon^u_{N},\epsilon_{rd})$ (see text for the details). (b): The ratio of cosmic ray to magnetic pressure $\beta_{cr}$, the degree of anisotropy A and the wave energy  of the slab modes $\epsilon_N$ at  the steady state vs kinetic energy of CRs in Galactic halo. (c): the steady state wave growth rate, turbulence compressional rate and damping rate. Growth only happens if the growth rate is larger than the damping rate. For the compression by fast modes, compressional rate is another threshold. (d): the mean free paths, obtained in this paper and in LB06.}
  \label{halo}
\end{figure*}

In interstellar medium, compressible turbulence is subjected to various damping processes. In the case of ionized phase, the damping of fast modes varies with the wave pitch angle $\theta$ (YL02, 04, 08). The turbulence  truncation scale increases with the pitch angle.  Approaching to $90^\circ$, the truncation scale is the largest partially because of the field line wandering (YL04, 08, see fig.\ref{halo}{\it left}), where the mean free path of cosmic rays $\lambda$ can be smaller than the truncation scale $l_c$. In the mean time, at smaller pitch angles, there is less damping and the mean free path $\lambda$ is larger than the truncation scale. In the low $\beta$ medium,  the fast modes are the dominant source for the compression of magnetic fields and we focus on fast modes for the low $\beta$ Galactic halo. Using the compression by the fast modes at the mean free path, we obtain  a fairly large growth of the waves $\epsilon_0$  with amplitude larger than that allowed by the energy budget of the turbulence (see Fig.\ref{halo}a). Turbulence is thus damped at the maximum mean free path of CRs $\lambda_{max}=r_p/\epsilon^u_{N}$. Compression below this scale is in fact reduced and eq.\ref{lambdlslc} should be adopted to recalculate the amplitude of the wave. We show the results in Fig.\ref{halo}.   

\begin{figure*}
\includegraphics[width=0.33\textwidth,
  height=0.24\textheight]{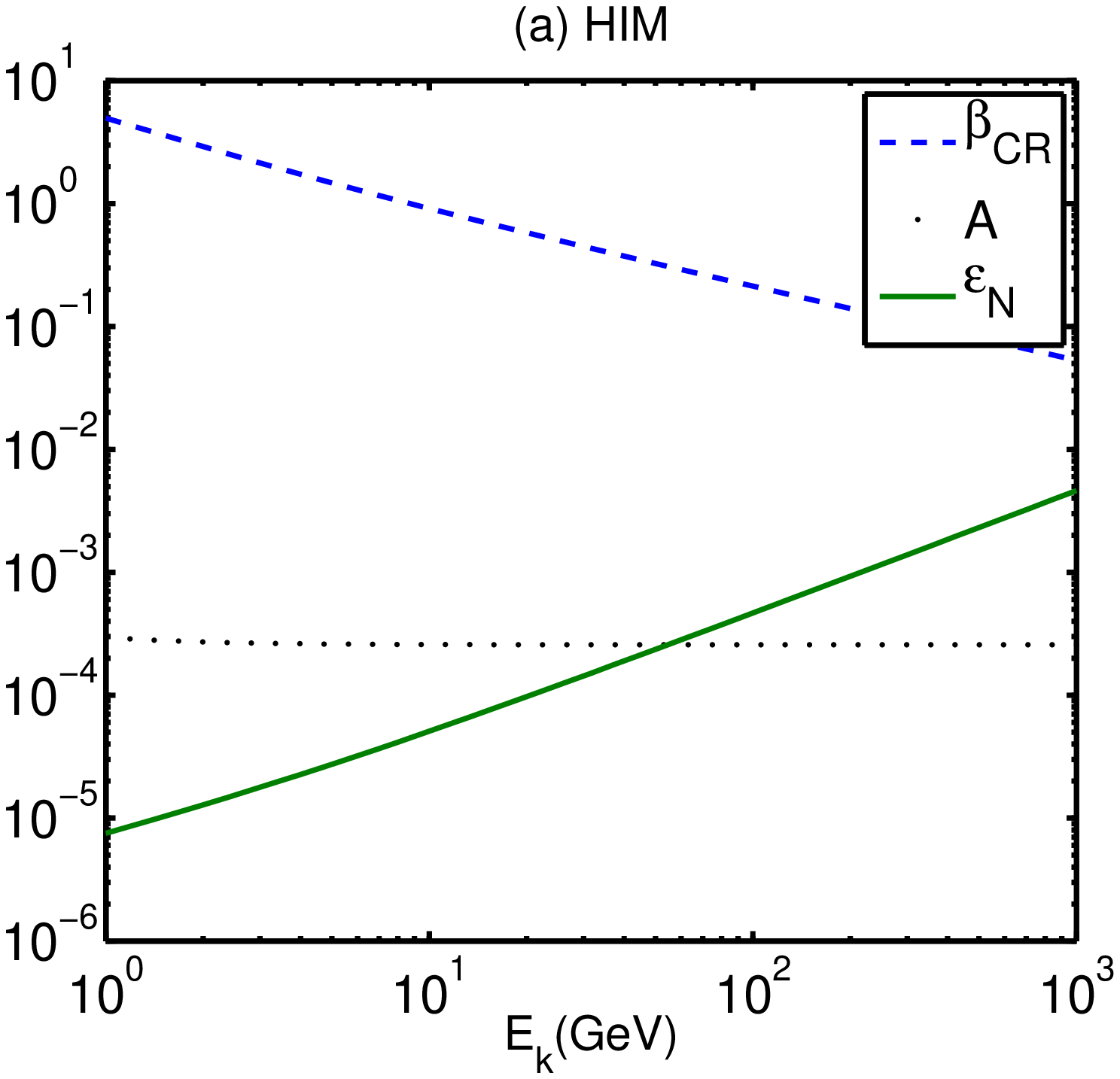}
  \includegraphics[width=0.33\textwidth,
  height=0.24\textheight]{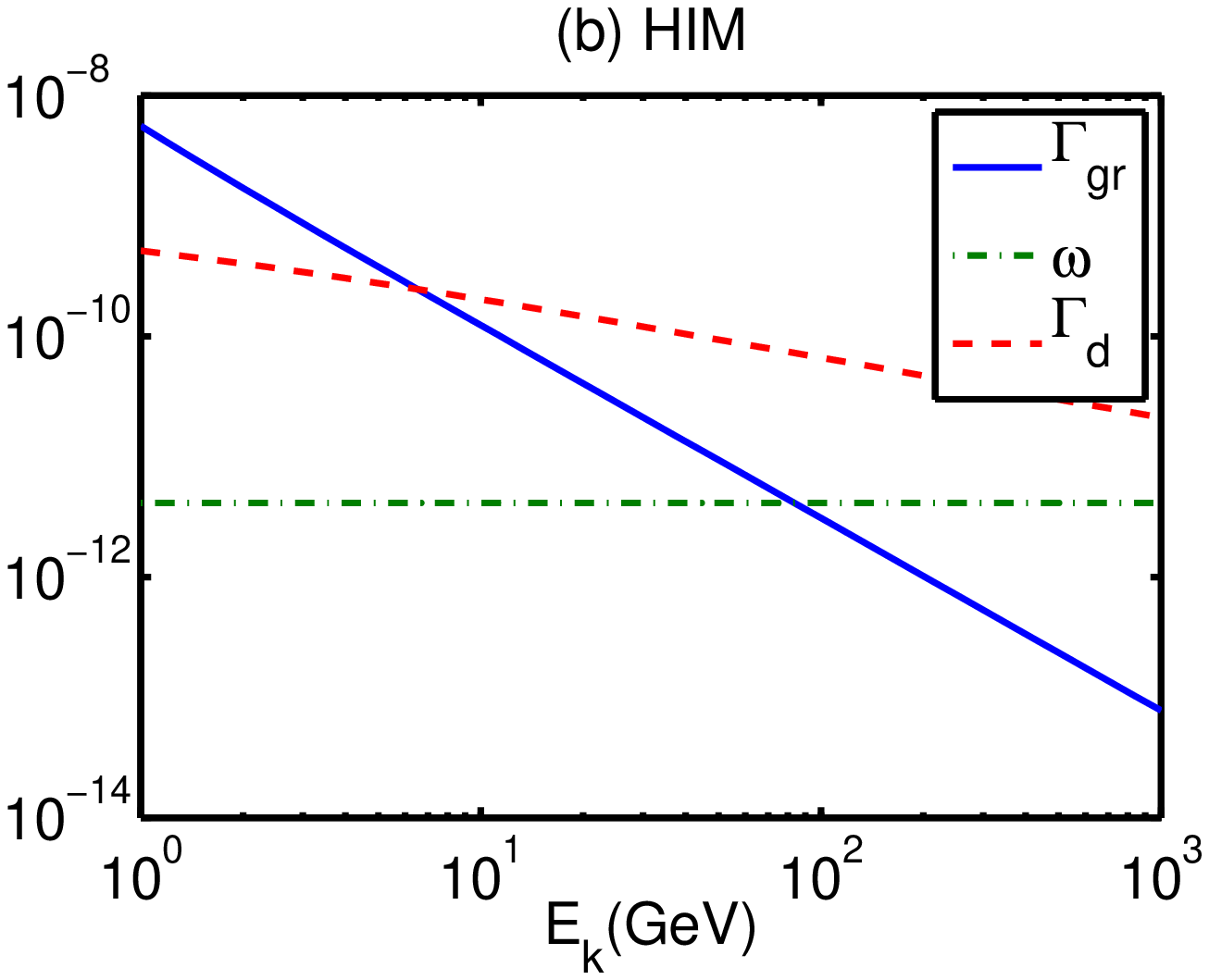}
   \includegraphics[width=0.33\textwidth,
  height=0.24\textheight]{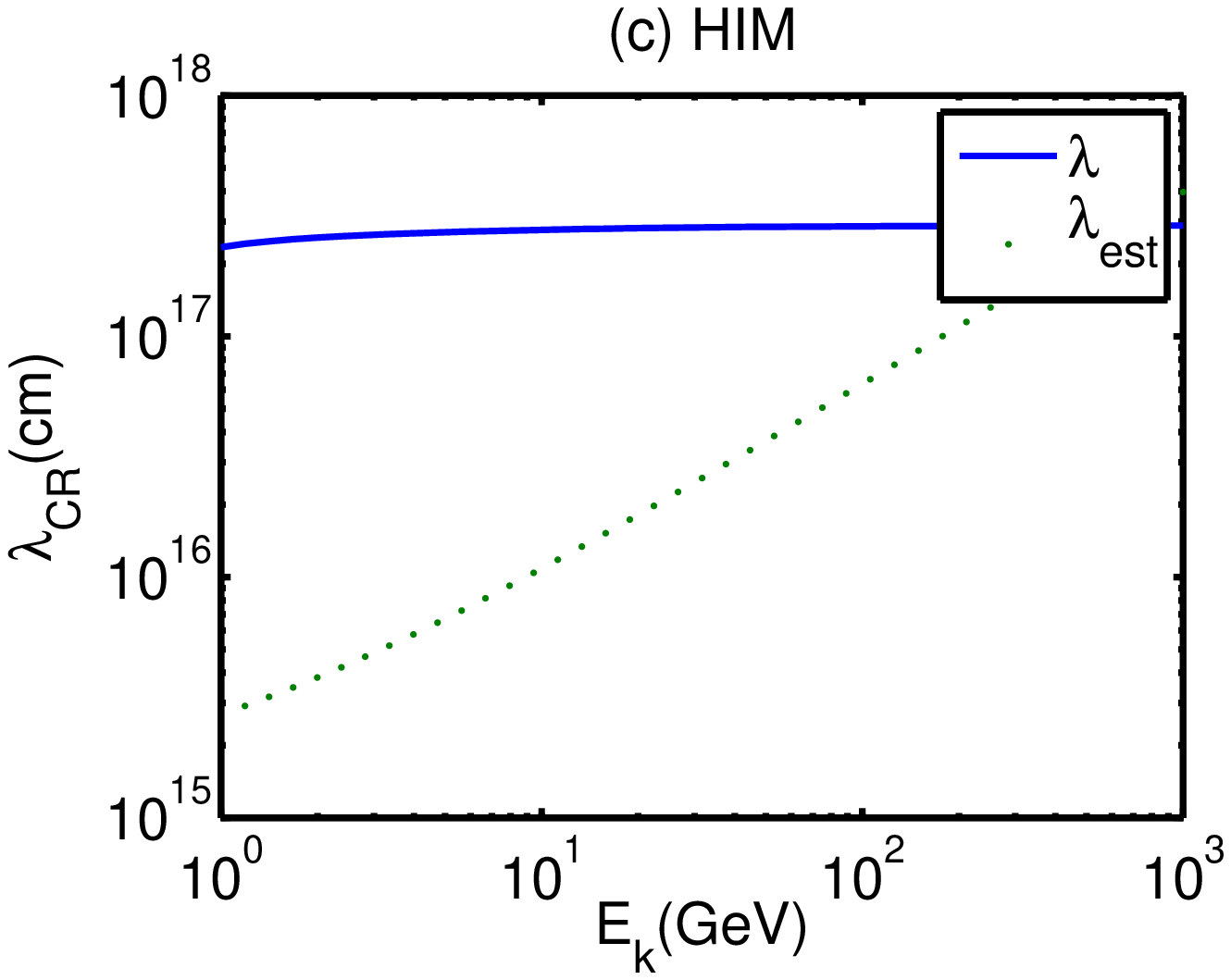}
  \caption{Same as Fig.\ref{halo}bcd, but for HIM.}
\label{HIM}
\end{figure*}

Another case we consider is hot ionized medium (HIM), which differs from halo in the sense $\beta>1$. In the high $\beta$ medium, the dominant compression originates from slow modes. In the collisionless medium, the damping scale of slow modes, the proton diffusion scale is quite large.  We use therefore Eq.(\ref{slowmodes_ls}) to calculate the wave amplitude. We check the consistency by comparing the final mean free path with the damping scale. The results are shown in Fig.\ref{HIM}. 

\subsubsection{Clusters of galaxies}

Observations of galaxy clusters show nonthermal component together with the thermal component in the intracluster medium. Turbulence may be induced by cluster of mergers and accretion of matter at the virial radius. Both observation and simulations suggest an appreciable amount of energy resides in turbulence. The typical velocity of turbulence at the injection scale is expected to be around 500-1000 km/s, which is highly super-Alfv\'enic because of high plasma $\beta$. In this case, the magnetic field correlation length $l_A$ could be smaller than the coulomb mean free path. The effective mean free path should be the minimum of the two if not accounting for further scattering of thermal particles due to plasma instabilities. 

\begin{figure*}
\includegraphics[width=0.33\textwidth,
  height=0.24\textheight]{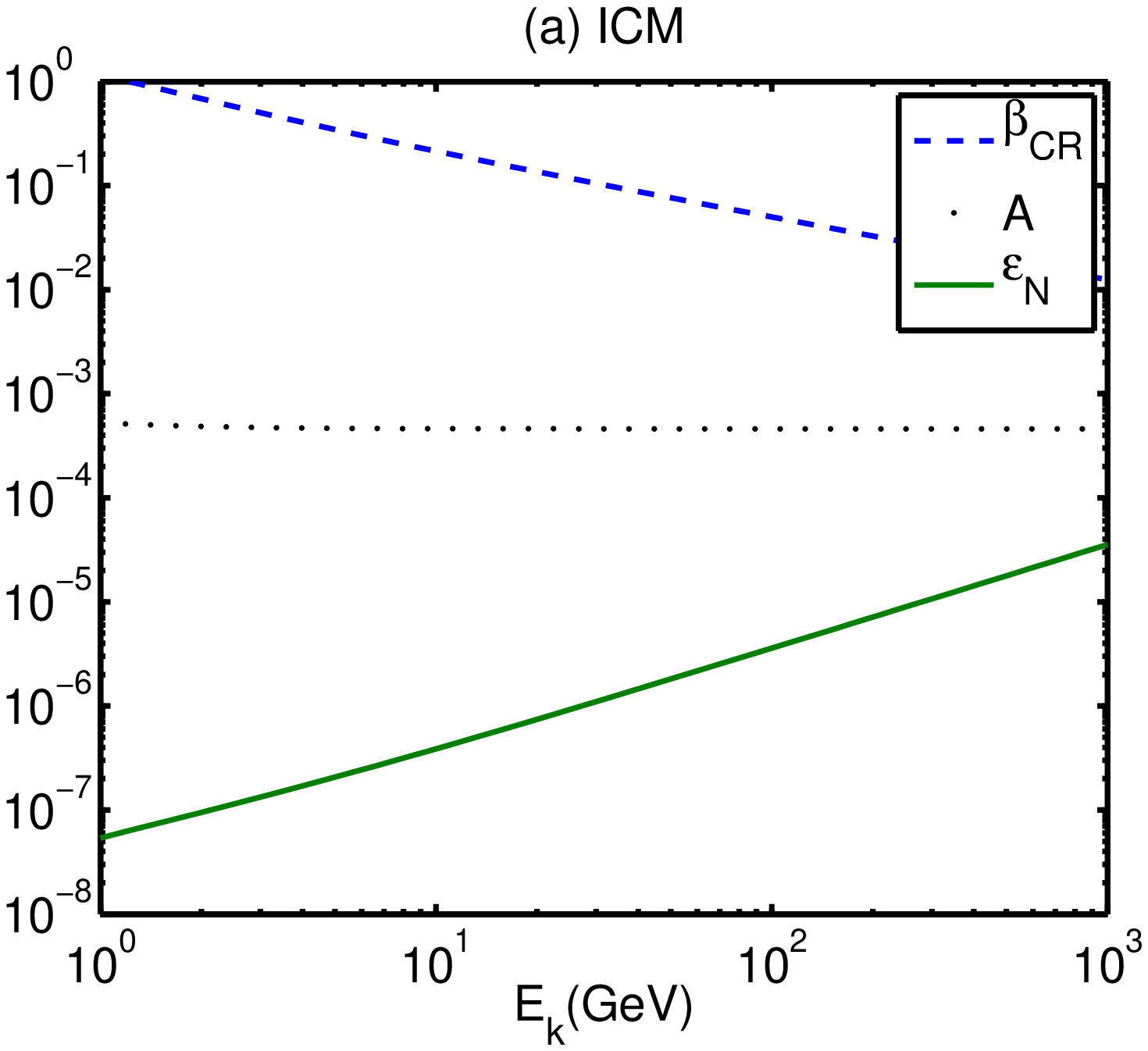}
  \includegraphics[width=0.33\textwidth,
  height=0.24\textheight]{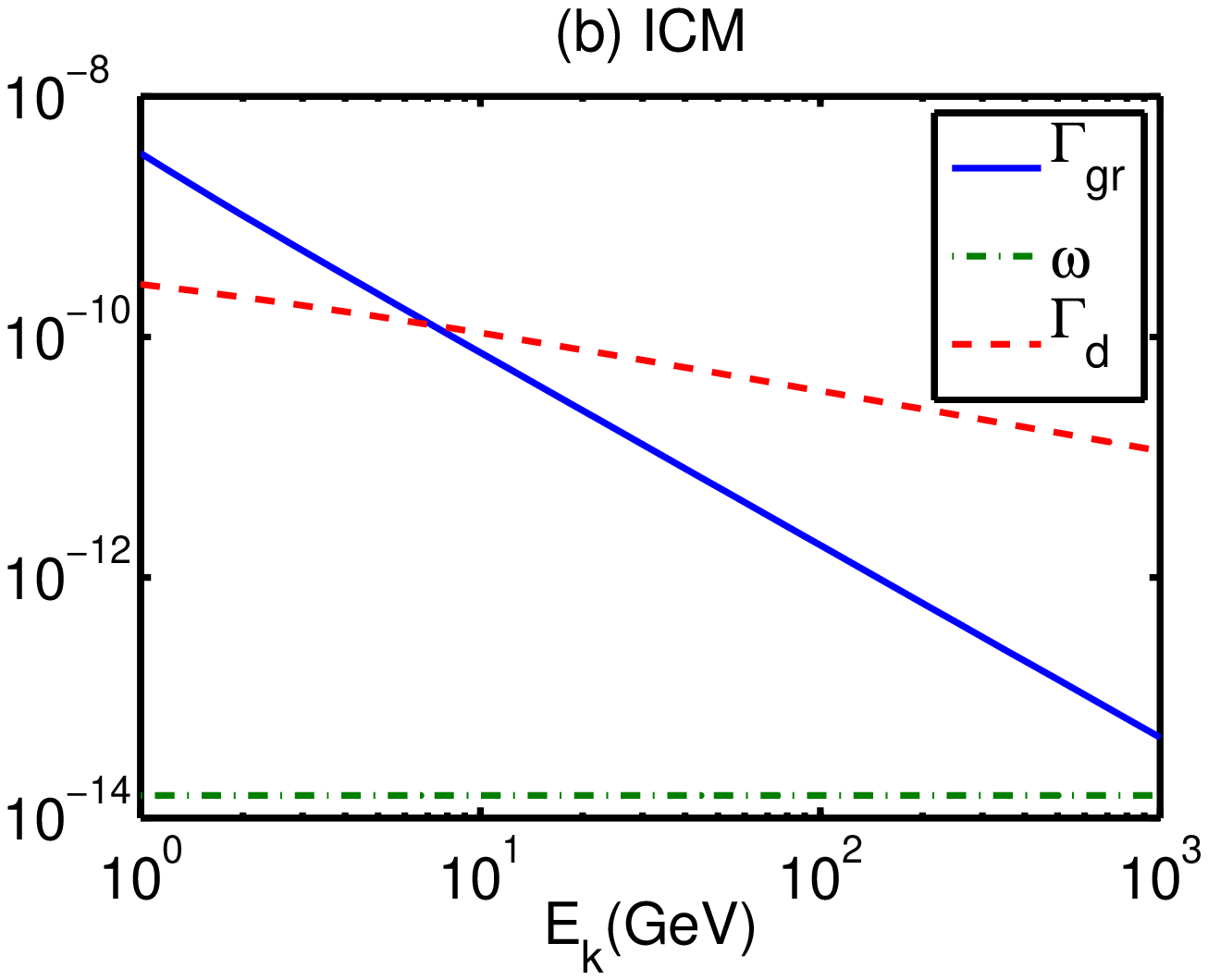}
   \includegraphics[width=0.33\textwidth,
  height=0.24\textheight]{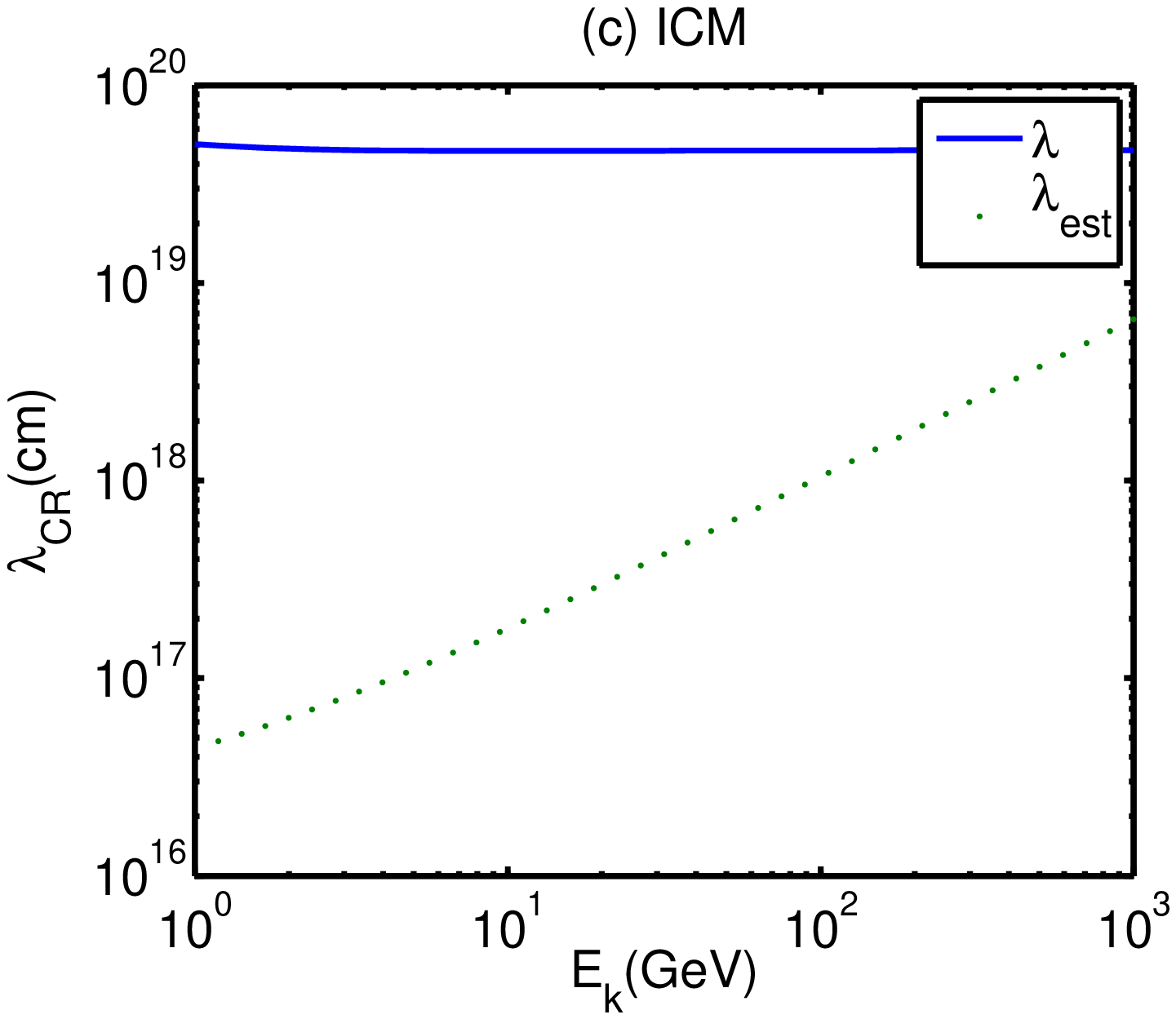}
    \caption{Same as Fig.\ref{halo}bcd, but for ICM.}
  \label{ICM}
\end{figure*}

Because of the similarity with the HIM, the results for intracluster medium (ICM) can be obtained with the similar procedure. They are shown in Figs.\ref{ICM}. Note that because of the plasma instabilities, the real thermal proton mean free path is less than the one we use. This means that the  mean free paths for CRs we obtain is an upper limit and practically, they can be even smaller.


\section{Discussion}

We demonstrate that the steady state amplitude of the slab wave increases with the CR energy. This is because among all the parameters, the growth rate changes ($\propto \gamma_L^{1-\alpha}$) most rapidly with the energy. And since the scattering rate linearly depends on the growth rate, the wave amplitude, inversely proportional to the scattering rate, increases with the energy of CRs (eqs.\ref{nu_est}, \ref{epsilon_est}). In the case that the growth is constrained by the energy budget of turbulence, the wave amplitude has a simple relation $\epsilon_N \propto \gamma_L^{\alpha-1} \propto k^{1-\alpha}$ on the energy (see Eq.\ref{energy}).


We show that the gyroresonance instability operates only for CRs of a limited energy range. On the low energy end, the  turbulence energy can limit the growth of the instability (as in the case of halo). On the high energy end, the growth is constrained by the feedback of the growing slab waves on the anisotropy of the CRs' distribution. The mechanism we discussed earlier (YL02,04,08), namely, CR scattering by large scale fast modes are, therefore, still dominant for higher energy CRs. 

Indeed the damping by background turbulence is the major constraint for the growth of the instability. In the environments where Alfv\'enic turbulence is weak or damped, the instability can extend to higher energies. For instance, in the regions where the ionization degree is low, the slab wave arising from the instability will protrude to a scale of $l_c^{4/3}/L^{1/3}$ (LB06). 


In the calculation on collisionless plasma, we used simply the coulomb mean free path. The effective viscosity in collisionless plasma should be less because of additional scattering of thermal particles caused by plasma instabilities. The consideration of this complex processes is beyond the scope of this paper. However, one thing we can be certain is that with the reduced viscosity, higher compressional efficiency can be achieved and the mean free path of CRs is further reduced and CRs are better confined, which is crucial for the acceleration of CRs.

Our results show that turbulence compression provides an additional channel for CR scattering and the feedback of CRs is important for the cascade of turbulence, confirming LB06's findings although our quantitative conclusions are very much different. In future MHD simulations, it is necessary to take into account the contribution from CRs.

\section{summary}

We have considered the gyroresonance instability of CRs in MHD turbulence. It arises from the compression from the compressible turbulence. The growth of the instability is self-adjusted due to the feedback of the slab wave on the anisotropy of CRs.  We find that

\begin{enumerate}
\item Compression by large scale turbulence results in anisotropic distribution of CRs, which induces gyroresonance instability. 

\item The growth of the instability is balanced by the feedback of the slab waves on the anisotropy of CRs. The high energy cutoff is determined by the balance between the growth rate and the damping of the wave by the large scale turbulence.

\item  The slab waves generated through gyroresonance instability limits the degree of the anisotropy of CRs, and is an efficient isotropization mechanism for CRs$< 100$GeV.

\item The feedback of the instability on the large scale turbulence should be taken into account. In some cases, turbulence cascade is truncated because of the drain of energy to the small scale waves. 

\end{enumerate}

\section*{Acknowledgments}
HY is supported by Arizona Prize Fellowship from U Arizona and 985 grant from Peking U. AL acknowledges the support by the NSF grant AST 0808118, NASA grant NNX09AH78G and the NSF-funded Center for Magnetic Self-Organization. 

\bibliography{ms3}

\begin{thebibliography}{39}
\expandafter\ifx\csname natexlab\endcsname\relax\def\natexlab#1{#1}\fi

\bibitem[{{Abramowitz} \& {Stegun}(1965)}]{Math_handbook}
{Abramowitz}, M., \& {Stegun}, I.~A. 1965, {Handbook of mathematical functions
  with formulas, graphs, and mathematical tables}, ed. {Abramowitz, M.~\&
  Stegun, I.~A.}

\bibitem[{{Barnes}(1966)}]{Barnes66}
{Barnes}, A. 1966, Physics of Fluids, 9, 1483

\bibitem[{{Beresnyak} \& {Lazarian}(2008)}]{BL08}
{Beresnyak}, A., \& {Lazarian}, A. 2008, \apj, 682, 1070

\bibitem[{{Beresnyak} \& {Lazarian}(2009)}]{BL09}
---. 2009, \apj, 702, 1190

\bibitem[{{Bieber} {et~al.}(1994){Bieber}, {Matthaeus}, {Smith}, {Wanner},
  {Kallenrode}, \& {Wibberenz}}]{Bieber94}
{Bieber}, J.~W., {Matthaeus}, W.~H., {Smith}, C.~W., {Wanner}, W.,
  {Kallenrode}, M., \& {Wibberenz}, G. 1994, \apj, 420, 294

\bibitem[{{Biskamp}(2003)}]{Biskampbook}
{Biskamp}, D. 2003, {Magnetohydrodynamic Turbulence}, ed. {Biskamp, D.}

\bibitem[{{Boldyrev}(2006)}]{Boldyrev}
{Boldyrev}, S. 2006, Physical Review Letters, 96, 115002

\bibitem[{{Brunetti} \& {Lazarian}(2007)}]{Brunetti_Laz}
{Brunetti}, G., \& {Lazarian}, A. 2007, \mnras, 378, 245

\bibitem[{{Cesarsky}(1980)}]{Cesarsky80}
{Cesarsky}, C.~J. 1980, \araa, 18, 289

\bibitem[{{Chandran}(2000)}]{Chandran00}
{Chandran}, B.~D.~G. 2000, Physical Review Letters, 85, 4656

\bibitem[{{Cho} \& {Lazarian}(2002)}]{CL02_PRL}
{Cho}, J., \& {Lazarian}, A. 2002, Physical Review Letters, 88, 245001

\bibitem[{{Cho} \& {Lazarian}(2003)}]{CL03}
---. 2003, \mnras, 345, 325

\bibitem[{{Cho} {et~al.}(2002){Cho}, {Lazarian}, \& {Vishniac}}]{CLV_incomp}
{Cho}, J., {Lazarian}, A., \& {Vishniac}, E.~T. 2002, \apj, 564, 291

\bibitem[{{Cho} {et~al.}(2003){Cho}, {Lazarian}, \& {Vishniac}}]{CLV_lecnotes}
{Cho}, J., {Lazarian}, A., \& {Vishniac}, E.~T. 2003, in Lecture Notes in
  Physics, Berlin Springer Verlag, Vol. 614, Turbulence and Magnetic Fields in
  Astrophysics, ed. E.~{Falgarone} \& T.~{Passot}, 56--98

\bibitem[{{Cho} \& {Vishniac}(2000)}]{CV00}
{Cho}, J., \& {Vishniac}, E.~T. 2000, \apj, 539, 273

\bibitem[{{Elmegreen} \& {Scalo}(2004)}]{ElmegreenScalo}
{Elmegreen}, B.~G., \& {Scalo}, J. 2004, \araa, 42, 211

\bibitem[{{Farmer} \& {Goldreich}(2004)}]{FG04}
{Farmer}, A.~J., \& {Goldreich}, P. 2004, \apj, 604, 671

\bibitem[{{Gary} \& {Tokar}(1985)}]{GaryTokar}
{Gary}, S.~P., \& {Tokar}, R.~L. 1985, \jgr, 90, 65

\bibitem[{{Gary} {et~al.}(1997){Gary}, {Wang}, {Winske}, \&
  {Fuselier}}]{Gary97_upper}
{Gary}, S.~P., {Wang}, J., {Winske}, D., \& {Fuselier}, S.~A. 1997, \jgr, 102,
  27159

\bibitem[{{Gogoberidze}(2007)}]{Gogo}
{Gogoberidze}, G. 2007, Physics of Plasmas, 14, 022304

\bibitem[{{Goldreich} \& {Sridhar}(1995)}]{GS95}
{Goldreich}, P., \& {Sridhar}, S. 1995, \apj, 438, 763

\bibitem[{{Kulsrud}(2005)}]{Kulsrudbook}
{Kulsrud}, R.~M. 2005, {Plasma physics for astrophysics}, ed. R.~M. {Kulsrud}

\bibitem[{{Lazarian} \& {Beresnyak}(2006)}]{LB06}
{Lazarian}, A., \& {Beresnyak}, A. 2006, \mnras, 373, 1195

\bibitem[{{Lazarian} \& {Pogosyan}(2000)}]{LP00}
{Lazarian}, A., \& {Pogosyan}, D. 2000, \apj, 537, 720

\bibitem[{{Lazarian} \& {Vishniac}(1999)}]{LV99}
{Lazarian}, A., \& {Vishniac}, E.~T. 1999, \apj, 517, 700

\bibitem[{{Lerche}(1967)}]{Lerche}
{Lerche}, I. 1967, \apj, 147, 689

\bibitem[{{Lerche} \& {Schlickeiser}(2001)}]{LercheSchlickeiser}
{Lerche}, I., \& {Schlickeiser}, R. 2001, \aap, 366, 1008

\bibitem[{{Lithwick} \& {Goldreich}(2001)}]{LG01}
{Lithwick}, Y., \& {Goldreich}, P. 2001, \apj, 562, 279

\bibitem[{Longair(1997)}]{Longairbook}
Longair, M.~S. 1997, High energy astrophysics. Volume 2: Stars, the galaxy and
  the interstellar medium (Cambridge U Press, 1994)

\bibitem[{{Maron} \& {Goldreich}(2001)}]{MG01}
{Maron}, J., \& {Goldreich}, P. 2001, \apj, 554, 1175

\bibitem[{{Melrose}(1974)}]{Melrose74}
{Melrose}, D.~B. 1974, \solphys, 37, 353

\bibitem[{{M{\"u}ller} \& {Biskamp}(2000)}]{MullerBisk}
{M{\"u}ller}, W., \& {Biskamp}, D. 2000, Physical Review Letters, 84, 475

\bibitem[{{Schlickeiser}(2002)}]{Schlickeiser02}
{Schlickeiser}, R. 2002, {Cosmic Ray Astrophysics}, ed. R.~{Schlickeiser}

\bibitem[{{Stanimirovi{\'c}} \& {Lazarian}(2001)}]{SL01}
{Stanimirovi{\'c}}, S., \& {Lazarian}, A. 2001, \apjl, 551, L53

\bibitem[{{Wentzel}(1974)}]{Wentzel74}
{Wentzel}, D.~G. 1974, \araa, 12, 71

\bibitem[{{Yan}(2009)}]{Yan09}
{Yan}, H. 2009, \mnras, 397, 1093

\bibitem[{{Yan} \& {Lazarian}(2002)}]{YL02}
{Yan}, H., \& {Lazarian}, A. 2002, Physical Review Letters, 89, B1102+

\bibitem[{{Yan} \& {Lazarian}(2004)}]{YL04}
---. 2004, \apj, 614, 757

\bibitem[{{Yan} \& {Lazarian}(2008)}]{YL08}
---. 2008, \apj, 673, 942

\end{thebibliography}
\bibliographystyle{apj}
\appendix

\section{Derivation of scattering efficiencies $S_\bot, S_\|$}

Define 
\be
D=\frac{\omega-kv_g}{v_\bot}\frac{\partial f_0}{\partial p_\bot}+k\frac{\partial f_0}{\partial p_\|}.
\ee


\bea
&&2e\Re<{\bf E} \cdot \Gamma>\nonumber\\
&=&2e\Re<{\bf E} \cdot \int d^3p{\bf v} f_1>\nonumber\\
&=&-\epsilon_\mp\Im\int d^3p \frac{8\pi e^2}{\omega}\frac{v^2_\bot}{kv_\|-\omega\mp \Omega}D\nonumber\\
&=&\frac{4}{|\omega|^2}\Im(\omega^2\omega^*K^\pm)\epsilon_\mp,
\label{Eperp}
\eea
where we applied Eq.(\ref{linearf},\ref{dielectric}) and the fact $\Re<A({\bf x},t)B({\bf x},t)>=1/2\Re[A^*({\bf k}, t)B({\bf k}, t)]$. We use K to express the result as it is directly associated with the growth rate (Eq.\ref{growth_def})

\bea
&&-\frac{2e}{c} \Re<\int d^3p{\bf v}_\bot\cdot \left[{\bf v}\times {\bf B}_1\right]f_1>\nonumber\\
&=&-2e\Re<\int d^3p {\bf v}_\bot\cdot \left[{\bf v}\times \frac{-i\nabla\times E}{\omega}\right] f_1>\nonumber\\
&=&-e\Re<\int d^3p kv_\|\frac{{\bf v_\bot\cdot E}}{\omega^*}  f_1>\nonumber\\
&=&-\epsilon_\mp\Im\int d^3p \frac{8\pi e^2}{|\omega|^2}\frac{Dv_\bot^2 kv_\| }{kv_\|-\omega\mp\Omega}\nonumber\\
&=&-\epsilon_\mp\Im\int d^3p \frac{8\pi e^2}{|\omega|^2}Dv_\bot^2\left(1+\frac{\omega\pm\Omega}{kv_\|-\omega\mp\Omega}\right)\nonumber\\
&=&-\epsilon_\mp\Im\int d^3p \frac{8\pi e^2}{|\omega|^2}v_\bot \frac{\partial f_0}{\partial p_\bot}\Im(\omega)+\frac{4}{|\omega|^2}\Im[\omega^2(\omega\pm \Omega)K^{\pm}]\epsilon_\mp
\label{Bscatt}
\eea
For a power law distribution $f_0\propto p^{-\alpha-2}, \, v_\bot \partial f_0/\partial p_\bot =-(\alpha+2)c p_\bot^2 p^{-\alpha-5}$. Insert it into the 2nd to the last term into Eq.(\ref{Bscatt}), we get
\bea
&&-\epsilon_\mp\int d^3p \frac{8\pi e^2}{|\omega|^2}v_\bot \frac{\partial f_0}{\partial p_\bot} \Im(\omega)\nonumber\\
&\simeq&\epsilon_\mp \frac{16\pi e^2}{(\alpha+1)(\alpha+3)|\omega|^2}\int d\left(\frac{p_\|}{p_{res}}\right) \frac{4n_{cr}(p>p_{res})}{\gamma_L m}\Im(\omega)\nonumber\\
&=&\frac{2(\alpha+2)}{(\alpha+1)(\alpha+3)} \int d\left(\frac{p_\|}{p_{res}}\right)\frac{4\omega_{pr}(p>p_{res})^2}{ |\omega|^2}\Im(\omega)\epsilon_\mp.
\eea

Therefore,
\bea
&&-\frac{2q}{c}<{\bf B}_1\cdot \int d^3p {\bf v}_\bot\times {\bf v}f_1>\\
&\simeq &\frac{2(\alpha+2)}{(\alpha+1)(\alpha+3)} \int d\left(\frac{p_\|}{p_{res}}\right)\omega_{pr}^2\Gamma\frac{4\epsilon_\mp}{|\omega|^2}+\Im\left[\omega^2(\omega\pm \Omega) K^\pm\right]\frac{4\epsilon_\mp}{|\omega|^2}.\nonumber
\label{parallel}
\eea 

From Eqs.(\ref{2ndmoment},\ref{Eperp},\ref{parallel}), we get
\bea
\frac{\partial W_\bot}{\partial t}&=&-4\sum\frac{2(\alpha+2)}{(\alpha+1)(\alpha+3)}\int d\left(\frac{k}{k_{res}}\right) \frac{\omega_{pr}^2}{|\omega|^2}\Gamma^\pm\epsilon_\mp+\Im\left[\frac{\omega^2}{|\omega|^2}(2i \gamma\pm\Omega_j)\epsilon_\mp K^\pm\right],\nonumber\\
\frac{\partial W_\|}{\partial t}&=& 4\sum\frac{2(\alpha+2)}{(\alpha+1)(\alpha+3)}\int d\left(\frac{k}{k_{res}}\right)\frac{\omega_{pr}^2}{|\omega|^2}\Gamma^\pm\epsilon_\mp+\Im\left[\frac{\omega^2}{|\omega|^2}(\omega_r\pm\Omega_j)\epsilon_\mp K^\pm\right],\eea
where $\sum$ is performed over the right moving CRs ('+') and left moving cosmic rays ('-'). In the above equations,
\bea
\omega^2&=&(\omega_r+i\gamma)^2=(\omega_r^2+\gamma^2)+2i\omega_r\Gamma,\nonumber\\
K^\pm&=&c^2/v_A^2(1-2i\Gamma^\pm/\omega_0).
\eea
Taking into account that $\gamma\ll \omega_r$ and $\omega_r\approx \omega_0$, we get

\bea
S_\bot(k)&=&\frac{\dot{W}_\bot(k)}{\dot{\epsilon_f}}=-2\left[\frac{2(\alpha+2)}{(\alpha+1)(\alpha+3)}\omega_{pr}^2+2k^2c^2\right]/\left(|\omega|^2+c^2k^2\right),\nonumber\\
&\simeq&-2\left[\frac{(\alpha+2)}{(\alpha+1)(\alpha+3)}\beta_{CR}(p>p_{res})+2\right],\nonumber\\
S_\|(k)&=&\frac{\dot{W}_\|(k)}{\dot{\epsilon_f}}=2\left[\frac{2(\alpha+2)}{(\alpha+1)(\alpha+3)}\omega_{pr}^2+k^2c^2\right]/\left(|\omega|^2+c^2k^2\right),\nonumber\\
&\simeq&2\left[\frac{(\alpha+2)}{(\alpha+1)(\alpha+3)}\beta_{CR}(p>p_{res})+1\right],
\eea
where we used the fact that $\partial \epsilon/\partial t=2\Gamma_{gr} (1+k^2c^2/|\omega|^2)(\epsilon_++\epsilon_-)$.

\label{lastpage}

\end{document}